\documentstyle[sprocl,amssymb]{article}

\begin{document}

\title{EXTRA DIMENSIONS AND SELF-ORGANIZING CRITICALITY
\footnote{Based on the talks given at ICTP Conference on "Physics Beyond Four Dimensions",
3-6 July 2000, Trieste, Italy, and at ISPM Workshop "Modern trends in Particles and Cosmology",
10-15 September 2000, Tbilisi, Georgia}}
\author{A.B.\ KOBAKHIDZE}

\address{HEP Division, Department of Physics, University of Helsinki and \\
Helsinki Institute of Physics, FIN-00014 Helsinki, Finland\\
E-mail: Archil.Kobakhidze@Helsinki.fi}

\maketitle

\abstracts{We discuss a possible explanation of the hierarchy problem within
the theories with spacetime dimensions higher than four. We show that the
presence of relatively (not hierarchycally) large extra dimensions can
significantly alter the evolution of the Higgs field VEV, driving it to an
infrared stable fixed point $\sim M_{W}$. Such a behaviour results in
self-organizing criticality and naturally explains gauge hierarchy without
any fine tuning of the parameters.}

\section{Introduction}

During the last few years a new explanations to the familiar hierarchy
between the fundamental high energy scales (say, the Planck scale M$%
_{Pl}\approx 10^{18}$ GeV) and the electroweak scale M$_{W}\approx 100$ GeV
have been proposed within the high dimensional theories \cite{1,2}. These
scenarios for solving the hierarchy problem are radically different from
those usually attributed to the supersymmetry or to the dynamical symmetry
breaking and explore the fact that the mass scales can be significantly
altered due to the geometry of extra space. The scenario of Refs. \cite{1}
utilize $\delta $ extra compact dimensions with large compactification radii
$r_{n}$ ($n=1,...,\delta $) in the factorizable, $M^{4}\times N^{\delta }$, (%
$4+\delta $)-dimensional spacetime where all known particles and
interactions (except the gravity) are localized on a four-dimensional
hypersurface $M^{4}$ (3-brane). Assuming then that the fundamental
high-dimensional scale M$_{*}$ is just an order of magnitude or so larger
than the electroweak scale M$_{W}$, the apparent weakness of gravity (or as
it is the same, the heaviness of the Placnk mass M$_{Pl}$) in the visible
four-dimensional world ($M^{4}$) is explained due to the large volume $%
V_{N^{\delta }}$ of the extra-dimensional submanifold $N^{\delta }$:
\begin{equation}
M_{Pl}^{2}=M_{*}^{\delta +2}V_{N^{\delta }}.  \label{1}
\end{equation}
The scenario of Ref. \cite{2} deals with a 5-dimensional non-factorizable AdS%
$_{5}$ space-time where two 3-branes located at the $S^{1}/Z_{2}$ orbifold
fixed points of the fifth compact dimension. Now the weakness of gravity in
the visible world 3-brane is explained without recourse to large extra
dimensions, but rather as a result of gravity localization on the hidden
3-brane. Gravity localization in such scenario occurs because the
five-dimensional Einstein's equations admit the solution for the space-time
metric with a scale factor (''warp factor'') which is a falling exponential
function of the distance along the extra dimension y perpendicular to the
branes\footnote{%
Another remarkable thing offered by such a solutions is the possibility to
reproduce the four-dimensional Newton's law in our universe even with
infinitely large (non-compact) extra dimensions \cite{3,4}.}:
\begin{equation}
ds^{2}=e^{-2k|y|}dx_{1+3}^{2}+dy^{2},  \label{2}
\end{equation}
Thus, graviton is essentially localized on the hidden brane with positive
tension which is located at $y=0$ fixed point of the $S^{1}/Z_{2}$ orbifold,
while the Standard Model particles are assumed to be restricted on the
visible brane with negative tension which is located at $y=\pi r_{c}$ ($r_{c}
$ is the size of extra dimension) orbifold fixed point. So, a hierarchically
small scale factor generated for the metric on the visible brane gives an
exponential hierarchy between the mass scales of the visible brane and the
fundamental mass scale $M_{*}$, after one appropriately rescales the fields
on the visible brane.

The crucial point of any successful solution of the hierarchy problem is a
stability of the hierarchy under the radiative corrections. For the
scenarios described above, the question of quantum stability is translated
to the problem of the stability of radii of extra dimensions. Indeed, while
the scenario of Refs. \cite{1} does eliminate the Planck/weak scale
hierarchy, it introduces a new the same order hierarchy between $\mu_{c}$ ($%
\mu_{c}=1/R_{c}$, where $R_{c}$ is a radius of extra dimensions) and M$_{W}$%
, $\frac{\mu _{c}}{M_{W}}\approx \frac{M_{W}}{M_{Pl}}$ \cite{1}. Thus the
stability of large extra dimensions remains as a critical question. The
radius stabilization is also crucial for the scenario of Ref. \cite{2},
although the hierarchy needed in this case is relatively small, $%
\mu_{c}/M_{Pl}\simeq 10^{-1}\div 10^{-2}$\footnote{%
Another deficiency of this model is the {\it ad hoc }fine tuning required
between the cosmological constants in the bulk and on the 3-branes, in order
to obtain the desired solution (\ref{2}). The stability of such fine tuning
seems also to be problematic.}. Several proposals to solve this problem have
been discussed in the literature (for some of them, see \cite{5}).

Here we would like to suggest an alternative mechanism for the solution of
gauge hierarchy problem in higher dimensional theories with relatively large
radii extra dimensions \cite{6}. We explore an old idea of "self-organizing
criticality" proposed in Refs. \cite{7}\footnote{%
Basically, the concept of self-organizing criticality where a certain
dynamical systems drive themselves to a critical state over the wide range
of length and time scales has been introduced in Ref.\cite{ad1} and
subsequently applied to a various systems starting from the multi-scale
structure of the natural world and ending by the economic and living systems
(see e.g. Ref.\cite{ad2} and references therein).}. The idea is the
following: The electroweak scale $M_{W}$ in the SM originates from the
spontaneous electroweak symmetry breaking and thus essentially determined by
the vacuum expectation value (VEV) of the scalar doublet field $%
<\phi>\approx 174$ GeV ($M_{W}\sim <\phi>$) which is an order parameter for
the electroweak phase transition. So the question why the electroweak scale
is so small compared with the fundamental high energy scales can be
reformulated: Why is the SM near the phase transition?, or: Why is the
system near criticality? A natural solution to the hierarchy problem then is
possible when the system is in a situation of self-organizing critically,
i.e. when it is near criticality not only for a particular tuning of theory
parameters (scalar mass, self-interaction coupling, etc... ) at high
energies, but for a wide range of them. Thus the self-organizing criticality
is possible if there are an infrared fixed-points (at least approximate) in
the evolution of a certain parameters and is closely related to large
anomalous dimensions \cite{7}. Indeed, let us consider the evolution
equation for the renormalized VEV $v_R$ of the scalar field:
\begin{equation}  \label{3}
\frac{dv^{2}_{R}(\mu)}{d(\ln\mu)}=Av^{2}_{R}(\mu),
\end{equation}
where $A$ is an anomalous dimension which generally depends on the running
parameters of the theory (self-interaction couplings, scalar-fermion Yukawa
couplings, gauge couplings, etc.)and $\mu$ is an energy scale. In the limit
of constant $A$ the solution to (\ref{3}) can be easily found:
\begin{equation}  \label{4}
\frac{v^{2}_{R}(M_{W})}{v^{2}_{R}(M_{Pl})}=\biggl (\frac{M_{W}}{M_{Pl}}%
\biggr )^{A}
\end{equation}
It is evident from (\ref{4}) that for $A=2$ the ratio $\frac{v^{2}_{R}(\mu)}{%
\mu^{2}}$ exhibits infrared stable fixed point behaviour, that, even for a
naturally expected large values of the VEV at high energies $%
v^{2}_{R}(M_{Pl})\simeq M_{Pl}$), could lead to a large desired hierarchy at
low energies.

However in four dimensions, $A\geq 2$ is highly undesirable, since anomalous
mass dimension $A$, being proportional to coupling constants, is usually $%
\ll 1$, unless some of the couplings (Higgs self-interacting coupling or/and
Yukawa couplings) are non-perturbative below the scale M$_{Pl}$, or there is
an unrealistically large number of degrees of freedom ensuring $A\geq 2$
\cite{7}. Needless to say, that it is very difficult (if ever possible) to
construct a realistic model obeying such conditions. In higher dimensional
theories, however, the situation is drastically changed. The point is that,
owing to the power-law (in contrast to the logarithmic in four dimensions)
evolution of the theory parameters, the Higgs vacuum expectation value
(VEV), while being of the order of M$_{Pl}$ at high energies, rapidly
decrease down to the infrared stable fixed point $\sim M_{W}$ even for the
small values of $A$, thus naturally inducing a large hierarchy even in the
case of SM with the ordinary number of colours and flavours.

\section{A toy model}

To be more quantitative, let us now a simplified example of the $SU(N)$%
-symmetric Higgs-Yukawa system with $N_{c}$ colours \cite{6}. Our starting
action in $D=4+\delta $ dimensions ($\delta $ is the number of extra compact
dimensions) is
\begin{eqnarray}
S_{\Lambda _{0}} &=&\int d^{4+\delta }[Z(\Lambda _{0})\partial _{\mu }\phi
^{+}\partial ^{\mu }\phi -\mu ^{2}(\Lambda _{0})\phi ^{+}\phi +\frac{1}{2}%
\lambda (\Lambda _{0})\left( \phi ^{+}\phi \right) ^{2}  \nonumber \\
&&+Z_{L}(\Lambda _{0})\overline{\psi }_{L}i\gamma _{\mu }\partial ^{\mu
}\psi _{L}+Z_{R}(\Lambda _{0})\overline{\psi }_{R}i\gamma _{\mu }\partial
^{\mu }\psi _{R}  \nonumber \\
&&+\left( h(\Lambda _{0})\overline{\psi }_{L}\phi \psi _{R}+h.c.\right) ],
\label{5}
\end{eqnarray}
where $\phi $ is an $N$-component complex scalar field, $\psi _{L}$ is an $N
$-component left-handed fermion field with $N_{c}$ colours and $\psi _{R} $
is a right--handed $SU(N/2)$-singlet fermion with $N_{c}$ colours again. $Z$%
, $Z_{L}$, and $Z_{R}$ in (\ref{1}) are the field renormalization factors
which we choose to be equal to 1 at the scale $\Lambda _{0}$. In the case of
$N=2$ and $N_{c}=3$ the action (\ref{1}) is just the SM action in the limit
of vanishing gauge couplings and Higgs-Yukawa couplings except for one type
of quarks.

Theory with action (\ref{1}) in higher dimensions ($\delta \neq 0$) is known
to be non--renormalizable, but it can be well defined by introducing an
ultraviolet cut-off $\Lambda _{0}$, which is natural to identify with the
fundamental Planck scale $M_{Pl}$. At low energies one can consistently
describe the theory using Wilsonian effective action approach \cite{8}. The
basic idea behind this approach is first to integrate out momentum modes
between a cut-off scale $\Lambda _{0}$ and lower energy scale $\Lambda $,
rather than to integrate over all momentum modes in one go. The remaining
integral from $\Lambda $ to zero may again be expressed as a partition
function, but the bare action $S_{\Lambda _{0}}$ (\ref{5}) is replaced by a
complicated effective action $S_{\Lambda }$ (Wilsonian effective action) and
the overall cut-off $\Lambda _{0}$ by the effective cut-off $\Lambda $, in
such a way that all physics, i.e. Green functions, are left invariant. The
difference in $S_{\Lambda }$ induced by the change of the cut-off is
determined integrating ''shell modes'' with momenta between $\Lambda $ and $%
\Lambda +\delta \Lambda $ and for an infinitesimal $\delta \Lambda $ becomes
a Gaussian path integral which can be exactly carried out. Thus, the scale
dependence of the Wilsonian effective action is given by the exact
functional differential equation
\begin{equation}  \label{6}
\Lambda \frac{\partial S_{\Lambda }}{\partial \Lambda }={\cal O}[S_{\Lambda
}]\ ,
\end{equation}
where ${\cal O}[S_{\Lambda }]$ is a non-linear operator acting on the
functional $S_{\Lambda }$. However, for the practical calculations it is
inevitable to approximate the evolution equation (\ref{6}). We have adopted
here so-called local potential approximation \cite{9} with a sharp cut-off
and have truncated the effective potential keeping only renormalizable terms
up to $\phi ^{4}$. In this approximation, defining the effective
renormalized four-dimensional VEV $\overline{v}$ self-interaction $\overline{%
\lambda}$ and Higgs-Yukawa $\overline{h}$ couplings through the
five-dimensional renormalized parameters $v_{R}$, $\lambda_{R}$ and $h_{R}$,
respectively, as\footnote{%
Here we assume that $\delta =D-4$ extra dimensions are compactified on a
circle of a fixed radius $R_{c}=\frac{1}{\mu _{c}}$. The factor $(2\pi
R_{c})^{\delta }$ is just the volume of extra space appeared in the
effective four-dimensional action after one integrates over the extra space.}%
:
\begin{eqnarray}  \label{7}
\overline{v}^{2}=(2\pi R_{c})^{\delta }v_{R}^{2},~~ \overline{\lambda }%
=(2\pi R_{c})^{-\delta }\lambda _{R},~~ \overline{h}=(2\pi R_{c})^{-\frac{%
\delta }{2}}h_{R},
\end{eqnarray}
we obtain eventually the following evolution equations (for more details see
Ref. \cite{6}):
\begin{equation}  \label{8}
\Lambda \frac{d\overline{v}^{2}}{d\Lambda }=\frac{K_{D}}{2}\left( \frac{2\pi
\Lambda }{\mu _{c}}\right) ^{\delta }\left[ -6\overline{\lambda }-\frac{2^{%
\frac{D}{2}+1}}{D}N_{c}\overline{h}^{2}+2^{\frac{D}{2}}N_{c}\frac{\overline{h%
}^{4}}{\overline{\lambda }}\right] \overline{v}^{2}\ ,
\end{equation}
\begin{equation}  \label{9}
\Lambda \frac{d\overline{\lambda }}{d\Lambda }=\frac{K_{D}}{2}\left( \frac{%
2\pi \Lambda }{\mu _{c}}\right) ^{\delta }\left[ (2N+8)\overline{\lambda }%
^{2}+\frac{2^{\frac{D}{2}+2}}{D}N_{c}\overline{h}^{2}\overline{\lambda }-2^{%
\frac{D}{2}}N_{c}\overline{h}^{4}\right] \ ,
\end{equation}
\begin{equation}  \label{10}
\Lambda \frac{d\overline{h}^{2}}{d\Lambda }=K_{D}\left( \frac{2\pi \Lambda }{%
\mu _{c}}\right) ^{\delta }\frac{2(N+1)+2^{\frac{D}{2}}N_{c}}{D}\overline{h}%
^{4}\,
\end{equation}
where $K_{D}=\frac{2^{1-D}}{\pi ^{-D/2}\Gamma (D/2)}$ is the $D$%
--dimensional angular integral. Note, that by taking $\delta =0$ the set of
Eqs. (8-10) correctly reproduces the familiar one-loop results of
perturbation theory in four dimensions. The crucial role of the extra
dimensions in solving the gauge hierarchy problem can be seen from Eqs.
(8-10) even without performing numerical calculations. Indeed, ignoring for
the moment the running of $\overline{\lambda }$ and $\overline{h}$, one
finds from (\ref{8})
\begin{equation}  \label{11}
\frac{\overline{v}(M_{W})}{\overline{v}(M_{Pl})}=\left( \frac{M_{W}}{\mu _{c}%
}\right) ^{\frac{\omega _{0}}{2}}\exp \left[ \frac{(2\pi )^{\delta }}{%
2\delta }\omega _{\delta }\left( 1-\left( \frac{M_{pl}}{\mu _{c}}\right)
^{\delta }\right) \right] \ ,
\end{equation}
where $\omega _{\delta }=\frac{K_{D}}{2}\left[ -6\overline{\lambda }-\frac{%
2^{\frac{D}{2}+1}}{D}N_{c}\overline{h}^{2}+2^{\frac{D}{2}}N_{c}\frac{%
\overline{h}^{4}}{\overline{\lambda }}\right]$. The exponential factor in (%
\ref{11}) can be naturally small in the case of extra dimensions ( $\delta
\neq 0$) even for small (but positive) values of $\omega _{\delta }$ ,
providing the desired hierarchy $\frac{\overline{v}(M_{W})}{\overline{v}
(M_{Pl})}\approx \frac{M_{W}}{M_{Pl}}$, while in four dimensions this ratio
is of the order of $O(1\div 10)$ unless $\omega _{0}=A_{0}\geq 2$, that can,
however, never be obtained in perturbation theory since for small couplings $%
A_{0}$ is proportional to these couplings \cite{7}, as already discussed
above.

Of course, the actual solutions of the set of Eqs. (8-10) is more
complicated, since the Yukawa and self-interaction couplings also exhibit
fast (power-law) running and the approximation of the constant $\overline{%
\lambda }$ and $\overline{h}$ is very crude. We have analyzed Eqs. (8-10)
numerically. The Yukawa coupling $\overline{h}$ rapidly decreases going down
in the energy region between $M_{Pl}$ and $\mu _{c}$ and drives to the
infrared stable fixed-point $\overline{h}=0$. If the Yukawa coupling
dominates over the $\overline{\lambda }$ ($\overline{h}^{2}\gg \overline{
\lambda }$) then $\overline{\lambda }$ at the same time increases for
smaller energies, until $\overline{\lambda }$ becomes large enough so that
the terms proportional to $\overline{\lambda } ^{2}$ and $\overline{h}^{2}%
\overline{\lambda }$ cancel the term proportional to $\overline{h}^{4}$ in (%
\ref{9}). Thus, $\overline{\lambda }$ approaches the infrared stable
fixed-point, $\overline{\lambda }\sim \overline{h}^{2}$. At the same time,
even starting with large initial values of $\overline{v} (M_{Pl})\lesssim
M_{Pl}$, $\overline{v}$ rapidly decreases and below the $\mu _{c}$ changes
very slowly. Thus, for certain $\mu _{c}$ and $\delta $ the mean value of
anomalous dimension $A_{\delta }$ can be equal to 2, which means that $\frac{%
v^{2}(\Lambda )}{\Lambda ^{2}}$ has an infrared stable quasi fixed-point.
Indeed, we have explicitly checked by solving numerically the system of Eqs.
(8-10), that the ratio $\frac{\overline{v}( \overline{v})}{\overline{v}%
(M_{Pl})}$ is actually stable under the variation of the scalar VEV at
Planck scale with $\overline{\lambda }(M_{Pl})$ and $\overline{h}(M_{Pl})$
fixed. For example, for $\overline{\lambda }(M_{Pl})=0.2$, $\overline{h}%
(M_{Pl})=3$, $\mu _{c}=10^{16.75}$GeV, $\delta =1$ we obtain for the average
value of the anomalous dimension between the scales $M_{Pl}$ and $\overline{v%
}$
\begin{equation}  \label{12}
\left\langle A\right\rangle \equiv (\ln \frac{\overline{v}}{M_{pl}}
)^{-1}\int_{\ln \frac{\overline{v}}{M_{pl}}}^{0}A_{\delta }(\Lambda )d(\ln
\frac{\Lambda }{M_{Pl}})\approx \frac{2}{1+0.03\ln \frac{M_{Pl}}{\overline{v}
(M_{Pl})}}.
\end{equation}
So, if $\overline{v}(M_{Pl})\approx M_{Pl}$, as it is naturally expected, $%
\left\langle A\right\rangle $ is close to 2, providing large stable
hierarchy $\frac{\overline{v}(\overline{v})}{\overline{v}(M_{Pl})}\approx
1.8\cdot 10^{-15}$. Thus, varying $\overline{v}(M_{Pl})$ by 10$\%$ around $%
10^{17}$GeV, we obtain $\left\langle A\right\rangle \approx 2.14\div 2.13$
and $\overline{v}(\overline{v})=157\div 190$ GeV. Furthermore, requiring
that $\overline{\lambda }$ and $\overline{h}$ are within the perturbative
regime ($\frac{\overline{\lambda }^{2}}{4\pi }<1$ and $\frac{\overline{h}^{2}%
}{4\pi }<1$) and that the relations (8-10) hold for the whole interval
between $M_{W}$ and $M_{Pl}$, our toy model predicts the upper bounds on the
physical masses of the scalar and fermion, respectively:
\begin{equation}  \label{13}
m_{S} \lesssim 73{\rm {GeV},~~ m_{F}\lesssim 100 {GeV}.}
\end{equation}

It should be stressed that our solution to the gauge hierarchy problem does
not require the extra dimensions to be large. In fact, the hierarchy $\frac{
\mu _{c}}{M_{Pl}}\sim 0.05\div 0.3$ is enough to get the desired values of $%
\overline{v}$ at low energies, even starting with naturally expected large
values of $\overline{v}$ at $M_{pl}$ ($\overline{v}(M_{pl})\sim M_{Pl}$).
Indeed, requiring that $\overline{v}(174$ GEV$)=174$ GeV and $\overline{%
\lambda }(174$ GEV$)=0.25$, $\overline{h}(174$ GEV$)=0.55$, we have obtained
numerically $\mu _{c}=10^{16.75}$, 10$^{17.30}$, 10$^{17.51}$ GeV for $%
\delta =1,2,$ and 3, respectively.

\section{Conclusions and outlook}

We have discussed here a possible explanation of the hierarchy problem
within the theories with spacetime dimensions higher than four. We have
shown that the presence of relatively (not hierarchycally) large extra
dimensions can significantly alter the evolution of the Higgs field VEV,
driving it to an infrared stable fixed point $\sim M_{W}$. Such a behaviour
results in self-organizing criticality and naturally explains gauge
hierarchy without any fine tuning of the parameters.

Let us conclude with the following comments. It is clear, that
more accurate calculations, related to an improved treatment of
thresholds and contributions beyond the LPA approximation and
truncation of the effective potential as well as higher loop
corrections quantitatively alter (perhaps quite significantly)
our predictions for particle masses in (\ref{13}), but
qualitatively the behaviour of the parameters seems to remain
unchanged, thus providing us with a the natural solution of the
gauge hierarchy problem as discussed above.

The idea discussed here can be also applied to explain other observed
hierarchies. The possible role of the fixed point solutions in generating
the fermion mass hierarchy within the high-dimensional theories have been
already discussed in\cite{9}. It is interesting to investigate whether or
not this mechanism can be applied for the solution to the cosmological
constant problem as well \cite{10}. Finally, while we have demonstrated our
mechanism for the solution of the gauge hierarchy problem on the simplified
model of Higgs-Yukawa interaction, we find no reason why it can not work in
the realistic models when a full set of the particles and forces of the SM
or its extensions will be included. Moreover, several examples of
nonsupersymmetric unification through extra dimensions have been presented
\cite{11}, so we believe that a unified model without the gauge hierarchy
problem and consistent with present experimental data can be constructed.

\section*{Acknowledgements}

This work was supported by the Academy of Finland under the Project No.
163394. ABK would like also to gratefully acknowledge the support from ICTP and the
Chancellor of University of Helsinki in attending the Conference
"Physics Beyond Four Dimensions", Trieste, 2000.

\end{document}